\documentstyle[twoside,fleqn,espcrc2]{article}

\newcommand{\AmS}{{\protect\the\textfont2
  A\kern-.1667em\lower.5ex\hbox{M}\kern-.125emS}}

\title{ Heavy-quark correlations in deep inelastic scattering}

\author{
        J. Smith \address{Institute for Theoretical Physics,
        State University of New York at Stony Brook,
        New York, 11794-3840, USA}
        \thanks {Partially supported under the contract NSF-09888.} 
        and
        B.W. Harris \address{Physics Department, Florida State University,
        Tallahassee, Florida, 32306-3016, USA}
        \thanks{Supported under the contract DOE-FG05-87ER40319.}
        }
        
\begin{document}

\begin{abstract}

We discuss some results for heavy quark correlations in next-to-leading 
order in deep inelastic electroproduction. 
\end{abstract}

\maketitle

\section{INTRODUCTION}

Order $\alpha_s$ QCD corrections to structure functions containing 
heavy quarks and to single heavy quark inclusive distributions 
in deep-inelastic electroproduction 
( ${\em i.\ \! e.\ \! }$, $\gamma^{\ast}(q) + P(p) \rightarrow Q(p_1) 
+ X$ where $X$ stands for any final hadronic 
state allowed by quantum-number conservation and $P(p)$ is a 
proton of momentum $p$ )
were recently published in \cite{LRSvN1} and \cite{LRSvN2}, 
respectively.  By combining the next-to-leading order (NLO) 
heavy quark structure functions with the corresponding 
light-quark structure functions \cite{ZvN}, the heavy quark 
content of the nucleon has been studied as a function of 
$Q^2 = -q^2$ and $x=Q^2/2p \cdot q$ \cite{LRSvN3}.  
Event rates for charm production integrated over bins 
in $x$ and $Q^2$ relevant to HERA data have been calculated in 
\cite{RSvN}.

To further the study of deep-inelastic electroproduction of 
heavy quarks we have recalculated the virtual-photon-parton cross 
sections of \cite{LRSvN1} in an exclusive fashion \cite{hs}.
This enables us to study the single and double differential 
distributions and correlations among all outgoing particles in the 
virtual-photon induced reaction 
$\gamma^{\ast} + P \rightarrow Q + \overline{Q} + X$ with $X=0$ 
or $1$ jet and to easily incorporate experimental cuts.  
By examining distributions and correlations 
that are trivial at lowest order (for example, the azimuthal angle between 
the heavy quark and the heavy antiquark) one directly tests the hard 
scattering cross section that is predicted by perturbative QCD and 
factorization. Therefore our results should give a clean 
test of NLO perturbative QCD\@.  
Nevertheless we remind the reader that this is a fixed order 
perturbative calculation and suffers from the same problems as all NLO 
calculations. There are regions in phase space where it will break down.  
For example, in the above mentioned azimuthal angle distribution, 
if one looks at the prediction too near the back-to-back 
configuration there will be an extra enhancement of the cross section due to 
multiple soft gluon emission which is not included in our NLO result.  

In this short report we discuss some 
details of the calculation and present interesting distributions.  
The transverse and longitudinal photon components are 
treated separately and the 
latest CTEQ3 parton densities \cite{CTEQ}, consistent with the 
newly released HERA data \cite{HERA}, 
are used in the kinematic regime appropriate for production of charm 
quarks at HERA. We make our predictions
at fixed values of $Q^2$ $(\geq 8.5 ({\rm GeV/c})^2)$ 
and $x$ $(\geq 4.2 \times 10^{-4})$.

We stress that here we only consider {\em extrinsic} heavy quark 
production, involving Bethe-Heitler and Compton  
production from massless partons.  For a discussion 
of {\em intrinsic} production, where the heavy quark is considered to 
be part of the proton's wavefunction, see Brodsky {\em et al}.\ \cite{int}.  
A variable flavor scheme which joins the extrinsic 
heavy flavor production picture at $\mu_{\rm phys} \ll m$ with a light mass 
parton density picture at $\mu_{\rm phys} \gg m$ has 
been discussed by Aivazis {\em et al}.\ \cite{var}.  
By comparing the fixed flavor scheme calculation 
of \cite{LRSvN1} with the variable flavor number scheme 
of \cite{var} it is concluded in \cite{comp} that the 
former yields the most stable and reliable results for 
$F_2(x,Q^2,m^2)$ in the threshold region 
( ${\em i.\ \! e.\ \! }, \, Q^2 \leq 10m^2$ 
where $m$ is the mass of the heavy quark). The contribution by W.L.
van Neerven to these Proceedings reviews the progress towards implementing
the variable flavour number scheme in NLO \cite{bmsmn}.

In our exclusive computation we use the subtraction method which is 
based on the replacement of divergent (collinear or soft) 
terms in the squared matrix elements by generalized plus distributions.  
This allows us to isolate the soft and collinear
poles within the framework of dimensional regularization without 
calculating all the phase space integrals 
in a spacetime dimension $n\ne 4$ as usually required in 
a traditional single particle inclusive computation.  
In this method the expressions for the squared matrix elements in the 
collinear limit appear in a factorized form, where poles in $n-4$ multiply 
splitting functions and lower order squared matrix elements. 
The cancellation of collinear singularities is then performed using the 
factorization theorem \cite{CSS}. 
The expressions for the squared matrix elements in the soft limit appear 
in a factorized form where poles in $n-4$ multiply lower 
order squared matrix elements.  The cancellation of soft singularities 
takes place upon adding the contributions from the renormalized 
virtual diagrams.
Since the final result is in four-dimensional space time, we can compute all 
relevant phase space integrations using standard Monte Carlo 
integration techniques \cite{lepage} and produce histograms 
for exclusive, semi-inclusive, or inclusive  quantities 
related to any of the outgoing particles.  We can also 
incorporate any reasonable 
set of experimental cuts.  Our computer code has no small phase space slice 
parameters and the parameters defining the generalized plus distributions 
may be tuned to give fast numerical convergence, 
which is an advantage of using this subtraction method.

\section{The $\gamma^{\ast}q$ channel}

Analysis of the partonic reaction
\begin{equation}
\label{quark}
\gamma^{\ast}(q) + q(k_1) \rightarrow q(k_2)+Q(p_1)+\bar{Q}(p_2) \, ,
\end{equation}
does not involve soft or virtual 
contributions so we can use it to most simply explain the method.
One particular graph out of the order $eg^2$ set is shown in fig.1.
We project this set of graphs on a particular photon polarization state $i$
and square the amplitude. The answer is written as $C_{i,q}M^q_i(3)$
where the coefficient absorbs overall factors.
\begin{figure}[tp]
\vspace{2.5cm}
\begin{picture}(7,7)
\includegraphics{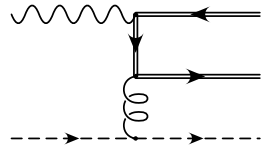}
\end{picture}
\caption{One order $eg^2$ diagram contributing to the amplitude for the 
reaction (1).  Additional graphs are obtained 
by reversing the arrows on the light-quark lines (dashed).}
\end{figure}
Now we write the photon-parton cross section as 
\begin{equation}
d \sigma_{i,q}^{(3)} = C_{i,q} M^q_i (3) d \Gamma_3\,. 
\end{equation}
where the 3 indicates a three-body final state.
To proceed we need some notation for the variables
$s = 2k_1\cdot q$, $ t_1 = t-m^2 = -2k_1\cdot p_2$, $u_1' = u_1-q^2 
= 2 q\cdot p_2$, $t' = -2k_1\cdot k_2$, $u'=q^2 - 2q\cdot k_2$ and
$s_5 = (p_1+p_2)^2 -q^2$. Then if we introduce $x = s_5/s$ and
$\cos^{-1} y = \vec q\cdot \vec k_2/|\vec q\cdot \vec k_2|$ we can write
$t' = -(1/2) s' (s'/s)(1-x)(1+y)$. This representation shows that 
$t'\rightarrow 0$ as $y\rightarrow -1$. This $t'$ divergence
yields the collinear divergence in $M^q_i(3)$, which is due to
the presence of the Feynman graph in fig.1. 
To handle this divergence we multiply and divide by $t'$ so that 
\begin{equation}
\label{fqi}
d \sigma_{i,q}^{(3)} = C_{i,q} 
f^q_i(x,y,\theta_1,\theta_2) \overline{d\Gamma_3}\, , 
\end{equation}
where
\begin{equation}
f^q_i(x,y,\theta_1,\theta_2) \equiv t^{\prime} \, M^q_i(3) \, , 
\end{equation}
is now finite as $y \rightarrow -1$ and 
\begin{equation}
\overline{d\Gamma_3} \equiv d\Gamma_3 / t^{\prime}.
\end{equation}
is divergent.
We next replace the factor
\begin{equation}
(1+ y)^{-1+\epsilon/2} = (1/(1+y))_\omega + \delta(1+y) (1/\epsilon
+ \ln \omega)\,,
\end{equation}
in the ($n$-dimensional) three-body phase space, where $0 < \omega < 2$.
This has the effect of splitting the $n$-dimensional cross section into 
two divergent parts. However it lets us take the limit $\epsilon 
\rightarrow 0$ and get back to 4-dimensions, with $\omega$
regularizing the collinear singularity.
Replacing the divergent factor $(1+y)^{-1+\epsilon/2}$ in 
$\overline{d\Gamma_3}$ one obtains the following decomposition:
(for more details see \cite{hs})
\begin{equation}
d \sigma_{i,q}^{(3)}=d \sigma^{(c-)}_{i,q}+d \sigma^{(f)}_{i,q},
\end{equation}
with 
\begin{eqnarray}
d \sigma^{(c-)}_{i,q} &=& - \frac{1}{\pi} C_{i,q} H N d\Gamma_2
(s^{\prime})^{\epsilon/2} \Big( \frac{s^{\prime}}{s} \Big)^{\epsilon/2}
\nonumber \\ & \times &
(1-x)^{\epsilon} (1-y)^{\epsilon/2} 
\delta (1+y) \Big( \frac{2}{\epsilon} + \ln \omega \Big) 
\nonumber \\ & \times &
dy \sin^{\epsilon} \theta_2 d\theta_2 f_i^q(x,y,\theta_1,\theta_2),
\end{eqnarray}
and
\begin{eqnarray}
d \sigma^{(f)}_{i,q} &=& - \frac{1}{\pi} C_{i,q} H N d\Gamma_2
(s^{\prime})^{\epsilon/2} \Big( \frac{s^{\prime}}{s} \Big)^{\epsilon/2}
\nonumber \\ & \times &
(1-x)^{\epsilon} (1-y)^{\epsilon/2} 
\Big( \frac{1}{1+y} \Big)_{\omega}
\nonumber \\ & \times & 
dy \sin^{\epsilon} \theta_2 d\theta_2 f_i^q(x,y,\theta_1,\theta_2).
\end{eqnarray}
The function $f$ satisfies
\begin{equation}
f^q_i(x,-1,\theta_1,\theta_2)=f^q_i(x,\theta_1)
+\tilde{f}^q_i(x,\theta_1,\theta_2) \, , 
\end{equation}
with
\begin{equation}
\int_0^{\pi} \tilde{f}^q_i(x,\theta_1,\theta_2) \sin^{\epsilon} 
\theta_2 d \theta_2 = 0.
\end{equation}
For reaction (1) we find that 
\begin{eqnarray}
\label{fqint}
f^q_i(x,\theta_1) &=& -128 \pi^2 \mu^{-2 \epsilon} 
\alpha_s^2 e^2 e^2_H N C_F 
\nonumber \\ & \times & 
\Big[ \frac{1+(1-x)^2+ \epsilon x^2/2}{x^2} \Big] 
\nonumber \\ & \times & 
(1+\epsilon/2)^{-1} 
B_{i,QED} (xk_1).
\end{eqnarray}
where $B_{i,QED}$ it the Born amplitude with $k_1$ replaced by $xk_1$.
The factor in the square brackets is the Altarelli-Parisi splitting function
in $n$ dimensions so we can define new (collinear free) cross sections
\begin{equation}
\label{quarkfinal}
d \hat{\sigma}_{i,q}^{(3)}=d\hat{\sigma}^{(c-)}_{i,q}+d\sigma^{(f)}_{i,q},
\end{equation}
with
\begin{eqnarray}
d \hat{\sigma}^{(c-)}_{i,q} &=& 8 C_{i,q} \alpha^2_s e^2 e_H^2 N C_F
B_{i,QED}(xk_1) d \Gamma_{2} 
\nonumber \\ & \times & 
\Big\{ 1+ \frac{1+(1-x)^2}{x^2} 
\Big[ \ln\frac{s^{\prime}}{\mu^2} 
+ \ln\frac{s^{\prime}}{s} 
\nonumber \\ & & 
+ \ln \frac{\omega}{2} + 2\ln(1-x)
\Big] \Big\} \, , 
\nonumber \\ && \\ 
d\sigma^{(f)}_{i,q} & = & - \Big( \frac{1}{16 \pi^2} \Big)^2 C_{i,q} 
\beta_5 f^q_i(x,y,\theta_1,\theta_2) 
\nonumber \\ & \times & 
\Big( \frac{1}{1+y} \Big)_{\omega} 
dx dy \sin \theta_1 d\theta_1d\theta_2. 
\end{eqnarray}
The finite functions $f^q_i (x,y,\theta_1,\theta_2)$ are available
in \cite{hs}.

As the quark channel only contains collinear singularities, 
we also use it to illustrate how the generalized plus 
distributions are implemented numerically 
and how the $\omega$ dependence disappears in the sum (\ref{quarkfinal}).  
To this end consider 
\begin{eqnarray}
\label{distex}
&d\sigma^{(f)}_{i,q} & \sim  \int_{-1}^1 dy f(y) \Big( \frac{1}{1+y} 
\Big)_{\omega} \nonumber \\
& = & \int_{-1}^{-1+\omega} dy f(y) \Big( \frac{1}{1+y} \Big)_{\omega} 
\nonumber \\ &&
  +   \int_{-1+\omega}^1 dy f(y) \Big( \frac{1}{1+y} \Big)_{\omega} 
\nonumber \\
& = & \int_{-1}^{-1+\omega} dy \frac{f(y)-f(-1)}{1+y} 
\nonumber \\ &&
  +   \int_{-1+\omega}^1 dy \frac{f(y)}{1+y}
\nonumber \\
& = & \int_{-1}^1 dy \frac{f(y)}{1+y} 
  -   \int_{-1}^{-1+\omega} dy \frac{f(-1)}{1+y}
\end{eqnarray}
where we have suppressed all indices and arguments of 
$f^q_i(x,y,\theta_1,\theta_2)$ other than $y$.
In the bottom line we see that the infinity encountered at the lower 
integration limit $y=-1$ is cancelled in the sum of the two integrals.  
In practice one can only reasonably take the lower limit to be 
$-1+\delta$ where $\delta \sim 10^{-7}$ in double 
precision {\scriptsize FORTRAN} before round off errors enter.  
None the less, the final 
results are stable with respect to the variation of $\delta$ in the 
range $10^{-5}$ to $10^{-7}$.  The upper 
limit of the second integral gives a contribution $f(-1) \ln \omega$ which 
cancels against the $\ln \omega$ term in $d \hat{\sigma}^{(c-)}_{i,q}$.  
The first integral in the bottom line is commonly called the ``event'' and 
has a positive definite weight.  The second integral plus the factorized 
collinear contribution is commonly called the ``counter-event'' and may have 
either positive or negative weight.  

The implementation of this procedure for the gluon channel 
is similar but more complicated due to the presence of 
both soft and collinear divergences.

\section{Results}

Recalling that the probability density is related to the momentum 
density via $f_{i/H} (\xi,\mu^2_f) = \xi \phi_{i/H}(\xi,\mu^2_f)$ we 
write the hadronic cross section as 
\begin{equation}
d \sigma_{\gamma^{\ast}H}(p) = \sum_i \int_{0}^{1} \frac{d \xi}{\xi} 
d \hat{\sigma}_i (\xi p) f_{i/H}(\xi,\mu^2_f).
\end{equation}
We now specialize to the case where $H$ is a proton, as in the case of 
HERA.  Using the relations
\begin{equation}
F_k = \frac{Q^2}{4 \pi^2 \alpha} \sigma_k,
\end{equation}
where $k=2,L$ with $\sigma_2 = \sigma_G + 3 \sigma_L / 2$, and relations 
for the scaling functions, we find 
\begin{eqnarray}
&&F_{k}(x,Q^2,m^2) = \frac{Q^2 \alpha_s(\mu^2)}{4\pi^2 m^2} 
\int_{\xi_{\rm min}}^1 \frac{d\xi}{\xi}  
\nonumber \\ && \qquad
\times  \,e_H^2 f_{g/P}(\xi,\mu^2) c^{(0)}_{k,g} \,  
\nonumber \\ &&
+ \frac{Q^2 \alpha_s^2(\mu^2)}{\pi m^2} \int_{\xi_{\rm min}}^1 
\frac{d\xi}{\xi} 
 \nonumber \\ &&
\times \Big\{ \,e_H^2 f_{g/P}(\xi,\mu^2) 
\Big( c^{(1)}_{k,g} + \bar c^{(1)}_{k,g} \ln \frac{\mu^2}{m^2} \Big) 
 \nonumber \\ &&
+ \sum_{i=q,\bar q} f_{i/P}(\xi,\mu^2) \Big[ e_H^2
\Big( c^{(1)}_{k,i} + \bar c^{(1)}_{k,i} \ln \frac{\mu^2}{m^2} \Big) 
 \nonumber \\ &&
+  e^2_i \, d^{(1)}_{k,i} + e_i\, e_H \, o^{(1)}_{k,i} \, 
\Big] \Big\} \, , \nonumber \\ &&
\end{eqnarray}
where $k = 2,L$.  We have set $\mu_f=\mu$ and shown the $\mu^2$ 
dependence of the running 
coupling $\alpha_s$ explicitly.  The lower boundary on the integration 
is given by $\xi_{\rm min} = x(4m^2+Q^2)/Q^2$. 
This formula yields the standard heavy quark 
structure functions $F_2(x,Q^2,m^2)$ and $F_L(x,Q^2,m^2)$ for 
electron proton scattering, and we will present results as differentials 
of these functions.  
From the formalism described in the previous section we are left with 
events of positive weight and counter-events of either positive or 
negative weight.  Our program  
outputs the final state four vectors of the event and counter-event 
together with the corresponding weight.  
We histogram these into bins to produce differential distributions.

\begin{figure}[tp]
\vspace{7cm}
\begin{picture}(7,7)
\includegraphics{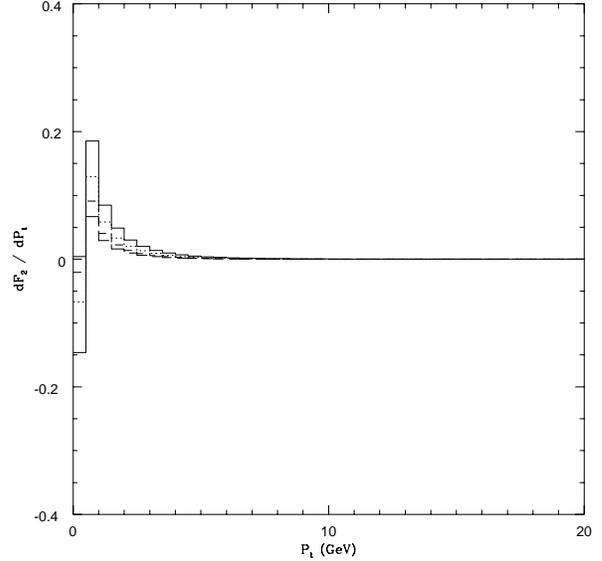}
\end{picture}
\caption{The distributions $dF_2(P_t)/dP_t$ for 
charm-anti\-charm pair production at fixed $Q^2=12 \, ({\rm GeV/c})^2$ 
with $x=$ $4.2 \times 10^{-4}$ (solid line), $8.5 \times 10^{-4}$ 
(dotted line), $1.6 \times 10^{-3}$ (short dashed line) and 
$2.7 \times 10^{-3}$ (long dashed line). }
\end{figure}

We start by considering the production of charm quarks.  We use
$m=m_c=1.5 \, {\rm GeV/c^2}$ and simply choose the factorization
(renormalization) scale as $\mu^2=Q^2+4(m_c^2+( P_t^c + P_t^{\bar{c}} )^2/4)$.
Note that there are many possible choices of scale as we have all 
components of the final four vectors.  Aside from the $P_t$ dependence, 
this choice reduces to $\mu^2=Q^2$ for electroproduction of massless quarks and
$\mu^2=4m_c^2$ for the photoproduction of charm quarks.  We introduce a $P_t$
dependence by adding in the average of the magnitude of the 
transverse momenta of the heavy quark and heavy antiquark.
As mentioned earlier we use the CTEQ3M parton densities \cite{CTEQ} in the
$\overline{\rm MS}$ scheme and the two loop $\alpha_s$ with
$\Lambda_{4} = 0.239 \, {\rm GeV}$.

The first distribution we present basically measures the transverse momentum 
of the additional jet which recoils against the heavy quark pair.
The $P_t$ distribution of the charm-anti\-charm pair is shown 
in fig. 2 where we plot $dF_2(x,Q^2,m_c^2,P_t) / dP_t$ 
as a function of $P_t$.  
\begin{figure}[tp]
\vspace{7cm}
\begin{picture}(7,7)
\includegraphics{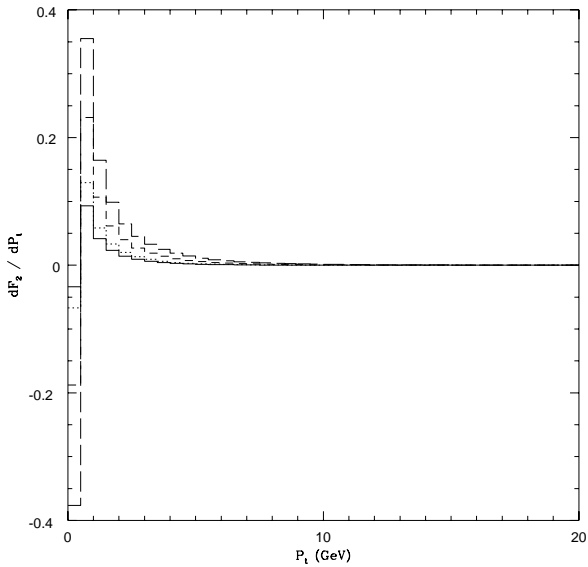}
\end{picture}
\caption{ The distributions $dF_2(P_t)/dP_t$ for 
charm-anti\-charm pair production at fixed 
$x=8.5 \times 10^{-4}$ and $Q^2=$ $8.5$ (solid line), 
$12$ (dotted line), $25$ (short dashed line), 
$50$ (long dashed line) all in units of $({\rm GeV/c})^2$.}
\end{figure}
The histograms are presented at fixed $Q^2$= 12 $({\rm  GeV}/c)^2$
for $x$ values of $4.2 \times  10^{-4}$, $8.5 \times  10^{-4}$, 
$1.6 \times  10^{-3}$ and $2.7 \times  10^{-3}$ respectively.
One sees that the $P_t$ distribution peaks at small 
$P_t$ and
has a negative contribution in the lowest bin.
This is a region where the dominant contribution is from counter-events 
so the weights can be negative.  
The results of this calculation require missing contributions 
from even higher order perturbation theory before this bin will have a 
positive weight. The depth of the negative bins 
is a function of $x$, $Q^2$, and the choice of scale.
Note that at larger $P_t$ the structure 
function is dominated by the contribution from the square of the 
bremsstrahlung graphs so the weights are positive.  
Figure 3 shows the corresponding results for fixed $x= 8.5 \times 10^{-4}$
plotted for the $Q^2$ values of 
$8.5$ $({\rm GeV}/c)^2$, $12$ $({\rm GeV}/c)^2$,
$25$ $({\rm GeV}/c)^2$ and $50$ $({\rm GeV}/c)^2$ respectively.
The distributions peak near small $P_t$ and are either small
or negative in the first bin.  The histograms with 
$Q^2 = 12 ( {\rm GeV} / c )^2 $ and $x=8.5 \times 10^{-4}$ 
(the dotted line) are the same in figs. 2 and 3.  We have also 
used the same scales on the axes so one can easily see that there is a greater 
change if we fix $x$ and vary $Q^2$ than if we fix $Q^2$ and vary $x$.  We 
will continue to use the same scale for all the pairs of later plots to 
simplify the comparison between them.

We now turn to the distributions in the azimuthal angle between the
outgoing charm quark and charm antiquark which we denote as $\Delta \phi$. 
This is the angle between the ${\bf P}_t$ vectors of the heavy quark-antiquark 
in the c.\ m.\ frame of the virtual-photon-hadron system.
Since we integrate over the azimuthal 
angle between the plane containing the incoming and outgoing leptons and the 
plane defined by the incoming parton and outgoing heavy quark 
(to limit our discussion to $F_2$ and $F_L$) we can only 
plot relative azimuthal correlations.  
In the Born approximation this distribution is a delta function at $\pi$, 
as their four momenta must balance.  Due to the radiation of the additional 
light mass parton, the distribution has a tail extending below $\pi$
and has a valley at $\pi$. 
The distributions become negative in the 
highest bins. This negative region is a general feature of all 
exclusive calculations.  
Figure 4 shows results for 
$dF_2(x,Q^2,m_c^2,\Delta \phi) / d(\Delta \phi)$
at the same values of fixed $Q^2$ and variable $x$ as chosen 
previously in fig. 2, while fig. 5 shows the 
results for fixed $x$ 
and variable $Q^2$ as chosen previously in fig. 3.  Note again 
that the dotted histograms are the same in figs. 4 and 5, and 
there is more variation for fixed $x$ and changing $Q^2$ than for 
fixed $Q^2$ and changing $x$.

In this report we have outlined the NLO calculation of the
virtual-photon-parton (Wilson) coefficient functions in the
fully differential production of heavy quarks plus one jet. More correlations
and checks are given in \cite{hs}. 
The computer program we have written for the exclusive calculation 
has the advantage that the four vectors of the heavy quark, heavy-antiquark 
and/or one additional light parton jet are produced for each event
and can be subjected to experimental cuts.  We have not done this in any 
of the plots shown here but the computer program is available and can be 
easily modified to incorporate acceptances of the detectors at HERA.
The program has been used to reanalyse the EMC data on charm production
in \cite{hsv}.
\begin{figure}[tp]
\vspace{7cm}
\begin{picture}(7,7)
\includegraphics{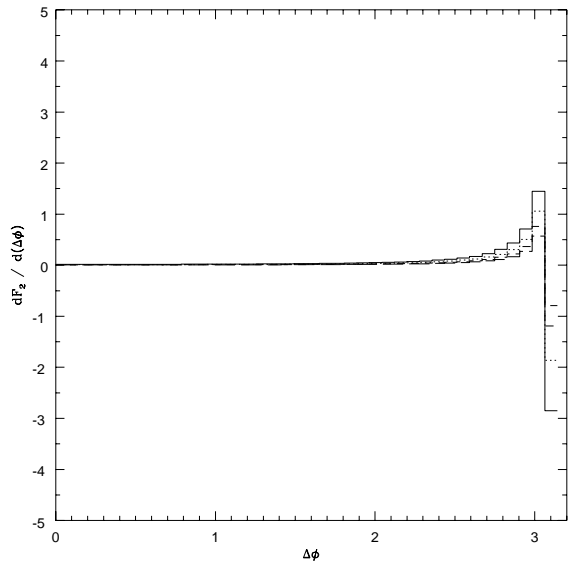}
\end{picture}
\caption{The distributions 
$dF_2(\Delta \phi)/d(\Delta \phi)$ for charm-anti\-charm 
pair production at the $x$ and $Q^2$ values given in Figure 2.} 
\end{figure}
\section{Conclusion}
In this report we have presented some plots which depend on
information from the four vector of the additional jet.
We showed the distributions in the transverse-momentum 
($P_t$) of the heavy quark antiquark pair and in the azimuthal angle
($\Delta \phi$) between the
${\bf P}_t$ vectors of the heavy quark and heavy antiquark. 
All quantities were predicted in the c.\ m.\ frame of the photon-proton 
system after integration over the azimuthal angle between the plane 
containing the incoming and outgoing lepton and the plane containing the 
incoming proton and outgoing heavy quark.  The results were presented as 
distributions in $F_2(x,Q^2,m^2)$  
at specific points in $x$, $Q^2$ and $m^2=m_c^2$.  
None of these distributions can be reproduced 
by any $K$-factor multiplication as the corresponding Born 
distributions are proportional to delta-functions.
In all cases the histograms of these 
distributions have negative bins. 
\begin{figure}[tp]
\vspace{7cm}
\begin{picture}(7,7)
\includegraphics{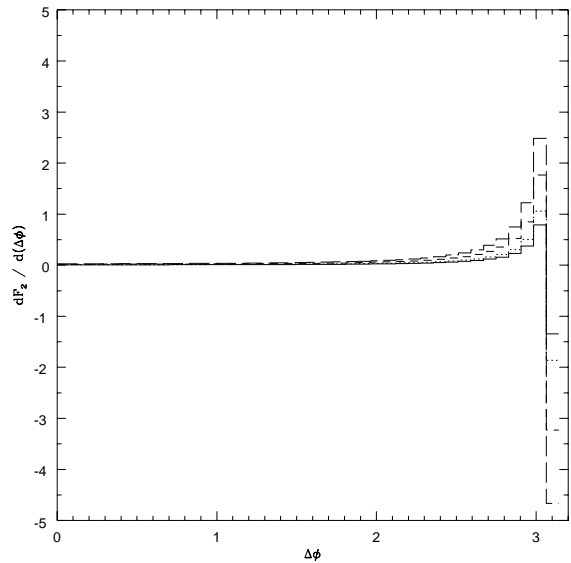}
\end{picture}
\caption{The same distributions at the $x$ and $Q^2$ values given in 
Figure 3.}
\end{figure}
These are regions where the
NLO calculation is not sufficient and a NNLO order
calculation (or some form of resummation) should be made.
A general statement about the magnitude of the NLO contribution 
compared to the LO one is difficult to make as the size and sign of 
the corrections may vary strongly between different regions of phase space.  
However, we see that all plots have larger $Q^2$ variation at fixed $x$ 
as comparied to varying $x$ at fixed $Q^2$.  
By varying the renormalization (factorization) scale we observed that 
the distributions presented here changed in normalization but not in shape. 
In our study we also calculated averages of various quantities and 
found the typical variation between central and extreme scale choices 
of 5 percent for charm production.

\end{document}